\documentclass[%
reprint,
superscriptaddress,
amsmath,amssymb,
aps,
prb,
]{revtex4-2}

\usepackage{booktabs}
\usepackage{multirow}
\usepackage{color}
\usepackage{graphicx}% Include figure files
\usepackage{dcolumn}% Align table columns on decimal point
\usepackage{bm}% bold math
\usepackage{hyperref}% add hypertext capabilities
\usepackage[mathlines]{lineno}% Enable numbering of text and display math
\usepackage{tabularx}

\begin{document}
	
%	\title{Floquet-engineered Valley Topology with Anisotropic Response based on Orthorhombic WSe$_2$ \& Janus-WSeTe}
	
	\title{Floquet-engineered Valley Topology with Anisotropic Response in 1T$^\prime$-WSe$_2$ and Janus WSeTe monolayers}
        
	\author{Zhe Li}%
	\email{lizhe21@iphy.ac.cn}
	\affiliation{%
		Beijing National Research Center for Condensed Matter Physics, and Institute of Physics, Chinese Academy of Sciences, Beijing 100190, China
	}%

	\author{Haijun Cao}
	\affiliation{%
		Beijing National Research Center for Condensed Matter Physics, and Institute of Physics, Chinese Academy of Sciences, Beijing 100190, China
	}%
	
	\affiliation{%
		University of Chinese Academy of Sciences, Beijing 100049, China
	}%

	\author{Lijuan Li}
	\affiliation{%
		Centre for Quantum Physics, Key Laboratory of Advanced Optoelectronic Quantum Architecture and Measurement (Ministry of Education), School of Physics, Beijing Institute of Technology, Beijing 100081, China
	}%

    \author{Huixia Fu}
	\affiliation{%
        Department of Physics and Center of Quantum Materials and Devices, College of Physics, Chongqing University, Chongqing 401331, China
	}%

    \affiliation{%
        Chongqing Key Laboratory for Strongly Coupled Physics, Chongqing University, Chongqing 401331, China
	}%

	\author{Mengxue Guan}
	\email{mxguan@bit.edu.cn}
	\affiliation{%
		Centre for Quantum Physics, Key Laboratory of Advanced Optoelectronic Quantum Architecture and Measurement (Ministry of Education), School of Physics, Beijing Institute of Technology, Beijing 100081, China
	}%

	\author{Sheng Meng}
	\email{smeng@iphy.ac.cn}
	\affiliation{%
		Beijing National Research Center for Condensed Matter Physics, and Institute of Physics, Chinese Academy of Sciences, Beijing 100190, China
	}%
	
	\affiliation{%
		University of Chinese Academy of Sciences, Beijing 100049, China
	}%
	
	\affiliation{%
		Songshan Lake Materials Laboratory, Dongguan, Guangdong 523808, China
	}%
	
	\date{\today}% It is always \today, today,
	
	\begin{abstract}
		Valley topology has emerged as a key concept for realizing new classes of quantum states. Here, we investigate Floquet-engineered topological phase transitions in anisotropic 1T$^\prime$-WSe$_2$ and its Janus derivative WSeTe monolayers, which exhibit valley-degenerate and valley-polarized characteristics, respectively. In 1T$^\prime$-WSe$_2$, a single topological-phase-transition (TPT) occurs from the quantum-spin-Hall state (QSH) to the quantum anomalous Hall (QAH) state, involving one spin channel at both valleys simultaneously. In contrast, Janus WSeTe undergoes a two-stage Floquet-driven TPT that occurs within a single valley and sequentially involves two spin components. The intermediate phase manifests as a valley-polarized QAH (vp-QAH) state with a finite valley Chern number, while the final phase evolves into a high-Chern-number QAH state with distinct valley gaps. Furthermore, an in-plane anisotropic response of the TPTs is predicted under oblique light incidence, reflecting the intrinsic low-symmetry nature of the lattice. These findings provide a comprehensive understanding of Floquet-engineered valley-based topological properties and offer guidance for designing light-controllable valleytronic and topological devices.
	\end{abstract}
	
	\maketitle
	
	\section{Introduction}
	% introduction
	The interplay between spin and valley degrees of freedom provides a versatile platform for understanding and manipulating topological quantum states \cite{xiao2012coupled,mak2014valley,xiao2007valley,pan2014valley,pan2015valley,zhang2015robust,zhang2018strong,vila2021valley,zhou2017valley,zou2020intrinsic,zhang2019converting,yang2025enhancement,li2024multimechanism,xue2024valley,vitale2018valleytronics,barman2024growth,hong2025designing,zhang2020abundant,li2025light,zhao2021valleytronic,li2022emergent,heissenbuttel2021valley,yao2024atomic,haldane1988model,kane2005quantum,bernevig2006quantum,hasan2010colloquium,qi2011topological,xu2013large}. Depending on the presence or absence of spatial inversion ($P$) and time-reversal ($T$) symmetries, distinct topological phases can emerge. When both $P$ and $T$ are preserved, band inversion driven by spin-orbit coupling (SOC) can yield the quantum spin Hall (QSH) state, characterized by helical edge channels and a vanishing total Chern number, $ C = 0$. Breaking $T$ while maintaining $P$ can transform the system into a quantum anomalous Hall (QAH) state with chiral edge modes and a nonzero Chern number. Alternatively, breaking $P$ while preserving $T$ gives rise to valley-distinct Berry curvature, leading to the quantum valley Hall (QVH) phase, where opposite Berry curvatures exist at symmetry-related valleys, yet the total $ C = 0$. Further breaking $T$ lifts the valley equivalence and generates valley-polarized QAH (vp-QAH) states, in which one valley dominates the topological response. Such spin-valley-coupled topologies have been investigated in buckled honeycomb group-IV Xenes, such as silicene, germanene, and stanene \cite{xiao2007valley,pan2014valley,pan2015valley,zhang2015robust,zhang2018strong,vila2021valley,zou2020intrinsic,zhang2019converting,yang2025enhancement,li2024multimechanism,xue2024valley,vitale2018valleytronics,barman2024growth,hong2025designing,li2025light}, and in 2H-phase transition-metal dichalcogenides (TMDs) \cite{xiao2012coupled,mak2014valley,heissenbuttel2021valley}, where SOC and broken inversion symmetry induce spin–valley locking and valley-selective optical transitions. Substrates or perpendicular electric fields can introduce intrinsic and extrinsic Rashba SOC \cite{pan2014valley,pan2015valley,li2024multimechanism,xue2024valley,hong2025designing}, while magnetic substrates further break $T$ and enable vp-QAH states \cite{pan2014valley,li2024multimechanism,xue2024valley,hong2025designing}. The coexistence of spin and valley degrees of freedom thus greatly enriches the family of accessible topological phases, providing a foundation for tunable valleytronics.\\

\begin{figure*}
		\centering
		\includegraphics[width=1\linewidth]{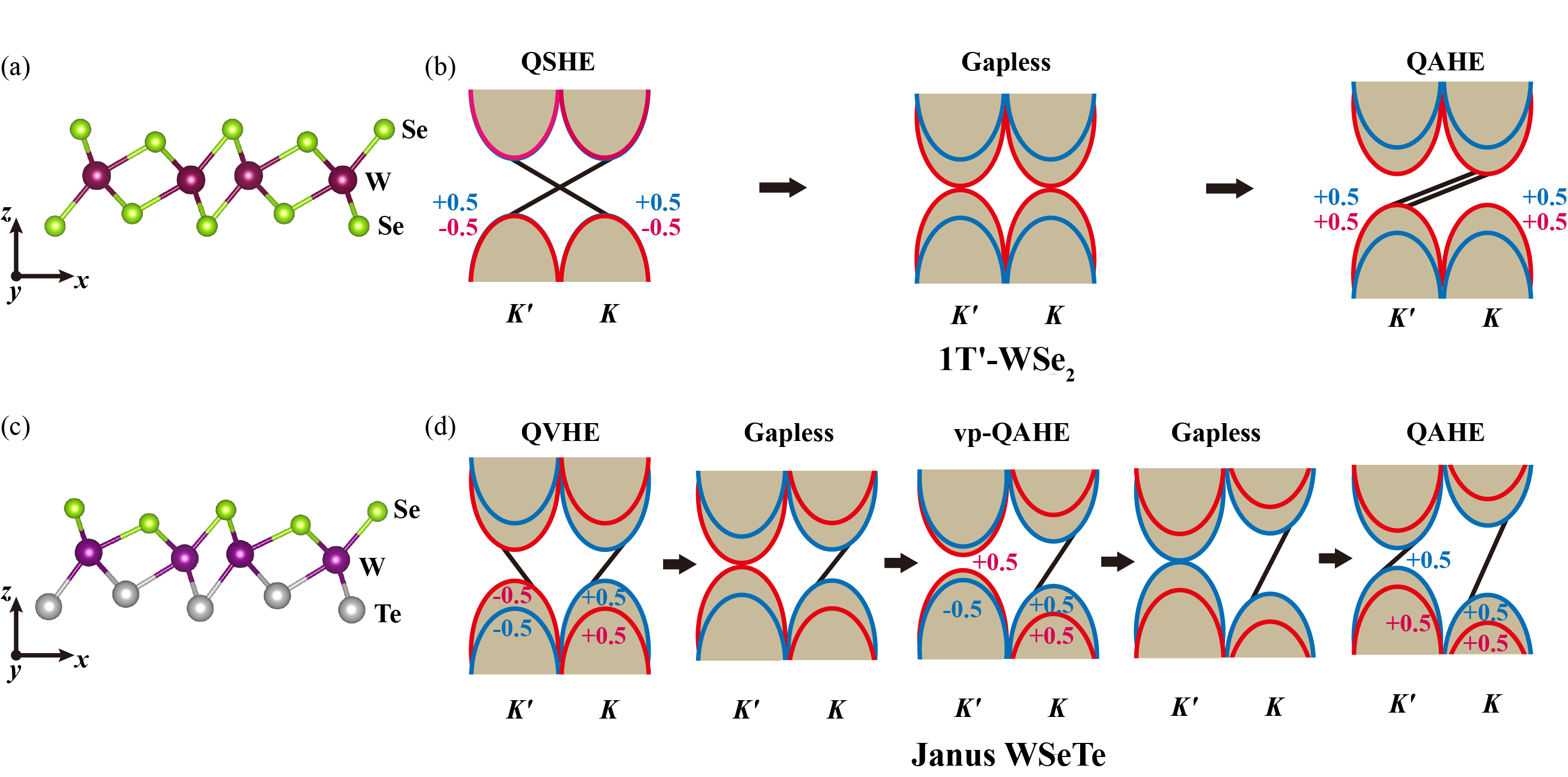}
		\caption{\textbf{Illustrations of different TPT processes for monolayer 1T$^\prime$-WSe$_2$ and Janus WSeTe.} (a) Side view structure of 1T$^\prime$-WSe$_2$, where the purple and light-green balls represent W and Se atoms, respectively. (b) Illustration of the TPT process in 1T$^\prime$-WSe$_2$ under irradiation of R-CPL. The light-brown region indicates the position of the 2D bulk states, with the red and blue edges corresponding to Chern-number contributions of --0.5 and +0.5, respectively. The black oblique line within the gap denotes the chiral edge state, where the positive or negative slope indicates positive or negative chirality. (c) and (d) are similar to (a) and (b), respectively, but depict the conditions for Janus WSeTe. The gray balls in (c) represent Te atoms.}
		\label{fig1:Illustration}
	\end{figure*}

	Beyond the hexagonal lattice framework, valley degrees of freedom can also emerge in other crystal systems with lower symmetry. For example, in square lattices, compounds such as LiFeSe \cite{li2020high}, MgFeP \cite{yao2024orbital}, and Fe$_2XY$ (where $X$, $Y$ = Cl, Br, I) \cite{guo2023possible,guo2024layer,guo2025first}, each exhibit two groups of valleys aligned with the $C_{4z}$ rotational symmetry. Moreover, orthorhombic 1T$^\prime$-phase TMDs combine strong SOC with pronounced in-plane anisotropy, providing an ideal platform for exploring coupled spin–valley topologies \cite{tang2017quantum,chen2018large,xu2018observation,li2020quantum}. Taking monolayer 1T$^\prime$-WSe$_2$ \cite{jain2013commentary,chen2018large,li2020quantum} as a typical representative, which possesses $P$ and is illustrated from the side view in Fig. \ref{fig1:Illustration}(a), we observe that it exhibits low symmetry (space group: 11, $P$21/m) while still demonstrating two-dimensional (2D) topologically nontrivial behavior at the two valleys along the $-Y$-$\Gamma$-$Y$ line, denoted as $K$ and $K'$ (see Fig. S2 of the Supplementary Materials\cite{supplementary}). The Janus counterpart WSeTe intrinsically breaks $P$ and introduces spontaneous valley polarization. The inherent anisotropy in their real- and momentum-space structures makes these systems particularly sensitive to external stimuli such as strain or polarized light, offering a promising route for the directional control of spin–valley topological states. \\
    
	Aiming to continuously manipulate the aforementioned spin-valley-coupled topologies, Floquet engineering emerges as the most convenient approach. It has been demonstrated to be a powerful tool for uncovering hidden topological properties in many traditional systems \cite{liu2023floquet,oka2009photovoltaic,bao2022light,zhan2024perspective}, leading to significant discoveries involving semimetal transitions in black phosphorus \cite{liu2018photoinduced}, Na$_3$Bi \cite{hubener2017creating}, and QAH induction in two-dimensional topological insulators \cite{li2025light,zhang2025quantum,qin2023light,li2025laser,zhu2025floquet,zhuang2025odd}. Researches have shown continuous phase transitions from high-order topology to QAH \cite{li2024floquet,tian2025quantized} and tunability of Chern numbers in materials like monolayer VSi$_2$N$_4$ \cite{zhan2023floquet} and thin-film MnBi$_2$Te$_4$ \cite{zhu2023floquet}. Notable experimental achievements include the observation of replica bands in materials such as black phosphorus \cite{zhou2023pseudospin,qiu2018ultrafast}, graphene \cite{choi2025observation,merboldt2025observation} and Bi$_2$Se$_3$ \cite{wang2013observation,mahmood2016selective}, along with the achievement of nearly quantized Hall conductance observed in circularly polarized light (CPL)-irradiated graphene \cite{mciver2020light}. Collectively, these advancements highlight the promising potential of Floquet band engineering in valley topology.
	
	This work delves into light-induced valley-based topological phase transitions (TPTs) in both 1T$^\prime$-WSe$_2$ and its Janus derivative WSeTe. In valley-degenerate 1T$^\prime$-WSe$_2$, a single TPT occurs from the QSH state \cite{kane2005quantum,bernevig2006quantum,hasan2010colloquium,qi2011topological} to the QAH state, driven by one spin channel. Conversely, the valley-polarized Janus WSeTe undergoes a two-stage TPT sequence, evolving from QVH to vp-QAH states and finally to the QAH phase through two successive band closures associated with opposite spin components at a single valley. Notably, both materials exhibit a pronounced anisotropic response of TPTs under oblique CPL incidence, reflecting their anisotropic valley distributions. Our findings facilitate precise manipulation of valley topology, providing valuable insights for future explorations of novel topological manifestations imposed by Floquet dynamics.
	
	\section{Methods}
	
	The ground states of monolayer 1T$^\prime$-WSe$_2$ and Janus WSeTe were obtained using first-principles computational methods operated by the Vienna ab initio Simulation Package (VASP) \cite{kresse1996efficient}, supplemented by the construction of a tight-binding Hamiltonian (TBH). A k-point mesh grid of 7$\times$9$\times$1 was employed for the relaxation and self-consistency calculations of the two systems. For the structural optimization, we applied the criterion that the Hellmann-Feynman force on each atom should be less than 0.001 eV/Å. Similarly, for electron energy convergence, we selected a threshold of 1.0$\times$10$^{-7}$ eV during the relaxation, self-consistency, and band structure calculations. The Perdew-Burke-Ernzerhof (PBE) functional was primarily used to determine their ground states \cite{perdew1996generalized}. In 1T$^\prime$-WSe$_2$ and Janus WSeTe, a vacuum layer of at least 15 Å was added to emulate two-dimensional behavior \cite{tkatchenko2009accurate}. Due to the monolayer structural nature, no van der Waals (vdW) corrections were applied. Spin-orbit coupling (SOC) effects were absent in relaxation step, but included in the self-consistency and band structure computation steps. VASPKIT \cite{wang2021vaspkit} was adopted for the post-processing of the band structures obtained from VASP, while PHONOPY \cite{togo2015first} was used to make adjustments to the lattice based on symmetry considerations. Besides, we employed the finite displacement method to acquire the phonon spectrum, which confirms the structural stability in these two systems.

	To obtain the excited states triggered by periodic optical fields, we first utilized the Wannier90 package to construct the TBHs for the two aforementioned systems, based on maximally localized Wannier functions \cite{mostofi2014updated,marzari1997maximally,souza2001maximally}. Following the acquisition of the ground-state TBHs, we employed a custom code to perform a Fourier transform of the Hamiltonian into reciprocal space, then proceeded by the application of Peierls substitution \cite{liu2023floquet}: $\textbf{\textit{k}}\ \rightarrow\textbf{\textit{k}}+\frac{e\textbf{\textit{A}}}{\hbar c}$. This was followed by an inverse Fourier transform to transition the Hamiltonian into real space. In our code, considering the high-frequency limit that was employed in this work, we integrated perturbations from higher-order side bands into the zeroth-order bands using a Magnus expansion approach \cite{liu2023floquet,bukov2015universal,blanes2009magnus}, extending up to ten orders:
	
	\begin{equation}
		{H_F}(\mathbf{k}) \approx H_{0,0}(\mathbf{k}) + \sum_{n=1}^{10} \frac{[H_{0,-n}(\mathbf{k}), H_{0,n}(\mathbf{k})]}{n \hbar \omega}
	\end{equation}

	Once we had derived the Floquet-engineered tight-binding Hamiltonian, we utilized the Green's function approach through the WannierTools package \cite{wu2018wanniertools} to investigate the topological characteristics of edge states, as well as to analyze Berry curvature distributions, spin-resolved band structures and the local density of states (LDOSs) that projected onto a certain edge.

		% Fig1

	\section{Structural Symmetries and Topological Phases}
	
	Monolayer 1T$^\prime$-WSe$_2$ is a 2D QSH insulator with valley degeneracy, where the two valleys exhibit $C_{2z}$ rotational symmetry in the 2D Brillouin zone (BZ). Constructing a Janus structure by replacing one atomic layer of Se with Te—resulting in monolayer WSeTe [Fig. \ref{fig1:Illustration}(c)]—breaks $P$ and reduces the space group from  $P$21/m to $P$m (No. 6), leading to valley polarization. In this transition, the QSH state is replaced by the QVH state, yielding a $Z_2 = 0$ and an opposite sign of nonzero Berry curvature in each valley (see Fig. S3 in Supplementary Materials \cite{supplementary}). This configuration results in both spins and valleys being split, while maintaining the same gap for each valley. We also verify the mechanical stability of these two systems by obtaining the phonon spectrum (see Fig. S4 \cite{supplementary}).

	%Fig2
	\begin{figure*}
		\centering
		\includegraphics[width=1\linewidth]{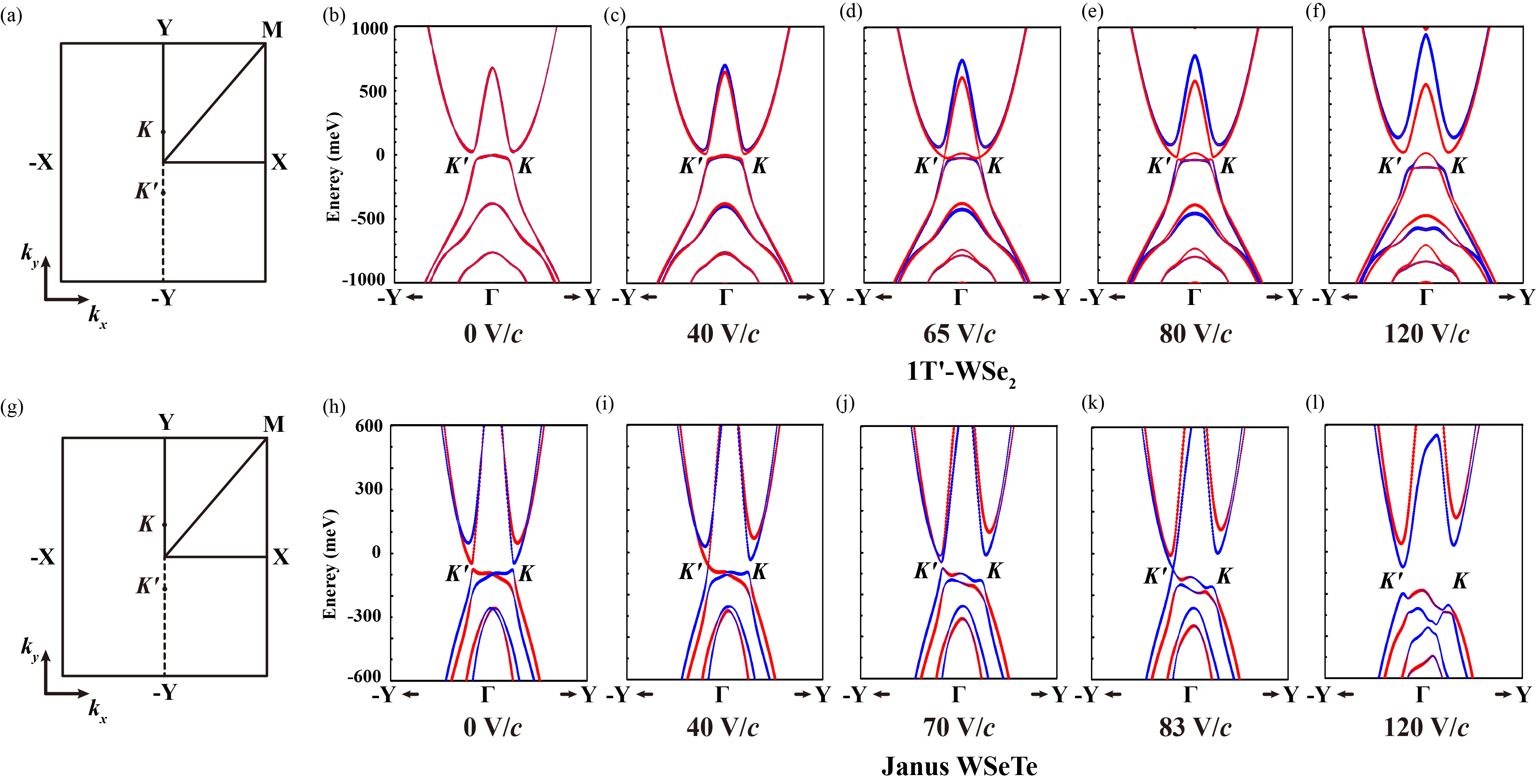}
		\caption{\textbf{Spin-resolved band evolutions near the two valleys of monolayer 1T$^\prime$-WSe$_2$ and Janus WSeTe, with the light frequency set to $\hbar\omega$ = 1.0 eV.} (a) Illustration of the BZ of 1T$^\prime$-WSe$_2$, where the high-symmetry lines $\Gamma$-$X$, $\Gamma$-$Y$ and $\Gamma$-$M$ are represented by solid black lines, while the line $\Gamma$-$-Y$ is indicated by a dashed black line. The positions of the two valleys are also marked. (b)-(f) sequentially depict zoomed-in spin-resolved band evolutions along the $K'$-$\Gamma$-$K$ path. Red and blue bubbles represent contributions from spin components “\textbf{1}” and “\textbf{2}”, respectively. The light intensity is set to $A_0$ = 0 V/$c$, 40 V/$c$, 65 V/$c$, 80 V/$c$ and 120 V/$c$. (g)-(l) are analogous to (a)-(f) but correspond to Janus WSeTe, with light intensities selected as 0 V/$c$, 40 V/$c$, 70 V/$c$, 83 V/$c$ and 120 V/$c$ sequentially.}
		\label{fig2:Spin-band}
	\end{figure*}
	
	Figures \ref{fig1:Illustration}(b) and \ref{fig1:Illustration}(d) illustrate the valley- and spin-resolved band evolutions of 1T$^\prime$-WSe$_2$ and Janus WSeTe under right-handed CPL (R-CPL), with spin components “\textbf{1}” and “\textbf{2}” colored by red and blue, respectively. In these one-pair-of-valley-based systems, each spin component devotes the Chern number of $\pm 1$, accompanied by spin degeneracy. For convenience, we propose that each spin-resolved valley contributes a Chern number of $\pm 0.5$ in 1T$^\prime$-WSe$_2$. This definition can also be generalized to valley-polarized conditions (Janus WSeTe). In 1T$^\prime$-WSe$_2$, both valleys simultaneously undergo band closure with a Chern number inversion from --0.5 to +0.5 for spin “\textbf{1}”. In contrast, Janus WSeTe initially has valley $K'$ with a Chern number of --0.5 for both spin components, while valley $K$ has a Chern number of +0.5. As the light intensity of CPL increases, the gap at valley $K$ broadens for both spin components, while spin “\textbf{1}” at valley $K'$ closes its gap and inverts its Chern number to +0.5, with the whole system transitioning into a vp-QAH state ($C$ = +1). Subsequently, spin “\textbf{2}” also closes its gap and inverts its Chern number to +0.5, ultimately leading to a QAH state with $C$ = +2, while maintaining distinct valley gaps.

	\section{Floquet-Driven Topological Phase Transitions}
	\subsection{Spin-resolved Bands \& Topological Edge States}
	
	The spin-resolved band-structure developments of 1T$^\prime$-WSe$_2$ and Janus WSeTe closely align with the illustrations presented in Figure \ref{fig1:Illustration}. Figures \ref{fig2:Spin-band}(a) to \ref{fig2:Spin-band}(f) depict the 2D BZ schematics and the spin-resolved bands of 1T$^\prime$-WSe$_2$ sequentially by setting the laser frequency ($\hbar\omega$) as 1.0 eV. The valley degeneracy leads to a simultaneous valley-gap-closing point at both valleys occurring at $A_0$ = 65 V/$c$, driven by spin “\textbf{1}” [see Fig. \ref{fig2:Spin-band}(d)]. This behavior shares underlying principles with that of group-IV Xenes \cite{li2025light}.
	
	% Fig3
	\begin{figure*}
		\centering
		\includegraphics[width=1\linewidth]{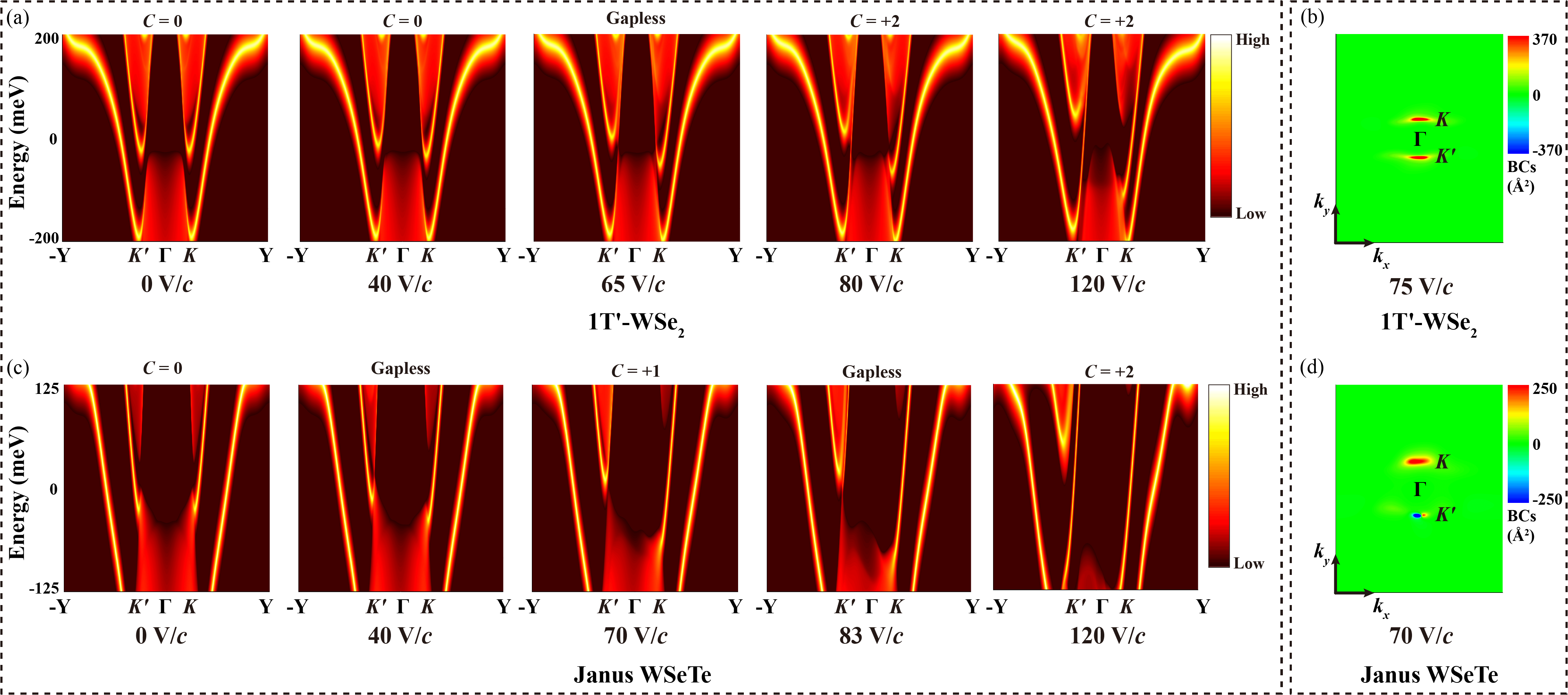}
		\caption{\textbf{Developments of LDOS patterns and selected Berry curvature distributions for monolayer 1T$^\prime$-WSe$_2$ and Janus WSeTe, with the light frequency set to $\hbar\omega$ = 1.0 eV.} (a) LDOS patterns projected onto the $X$ edge of 1T$^\prime$-WSe$_2$, with light intensities chosen as $A_0$ = 0 V/$c$, 40 V/$c$, 65 V/$c$, 80 V/$c$ and 120 V/$c$ corresponding to the panels from left to right. The color gradient transitions from black to red, yellow, then to white, indicating an increase in LDOS values. The topological state associated with each panel is indicated at the top. (b) Berry curvature distributions of 1T$^\prime$-WSe$_2$ within one 2D BZ. The color gradient from green to red (blue) represents the enhancement of Berry curvature with positive (negative) chirality. The light intensity is set to 70 V/$c$, corresponding to the QAH state. (c) and (d) are analogous to (a) and (b), respectively, but relate to Janus WSeTe. In subfigure (c), the light intensities are 0 V/$c$, 40 V/$c$, 70 V/$c$, 83 V/$c$ and 120 V/$c$. In subfigure (d), the light intensity is selected corresponding to the vp-QAH state.}
		\label{fig3:LDOSs}
	\end{figure*}

	In contrast, for Janus WSeTe, two distinct band closing points emerges at the $K'$ valley, serially governed by spins “\textbf{1}” and “\textbf{2}”, located at 40 V/$c$ and 83 V/$c$ respectively [Figs. \ref{fig2:Spin-band}(i) and \ref{fig2:Spin-band}(k), $\hbar\omega$ = 1.0 eV]. At $A_0$ = 0 V/$c$ [Fig. \ref{fig2:Spin-band}(h)], the two spin components exhibit symmetric splitting, contrasting with the two valleys at opposite momenta, in accordance with $P\sigma$, resembling characteristics of altermagnets \cite{vsmejkal2022emerging,zhou2024crystal,duan2025antiferroelectric,gu2025ferroelectric,han2025discovery,feng2025type,li2025floquet-am,huang2025light,zhu2025floquet,zhuang2025odd}. The introduction of CPL disrupts this relationship, yielding the aforementioned two-stage TPT process. Conversely, when the chirality of the light is flipped to left-handed CPL (L-CPL), not only does this reverse the gap-closing sequence of the two spin components (with spin “\textbf{2}” closing first), but it also shifts the gap-closing point to the $K$ valley, as summarized in Fig. S5 \cite{supplementary}. This finding reveals the intricate interplay between spin and valley components.
	
	The previously discussed relationship provides a richer landscape for valley topology. Figure \ref{fig3:LDOSs} presents the $X$-edge projected LDOS patterns for 1T$^\prime$-WSe$_2$ [Fig. \ref{fig3:LDOSs}(a)] and Janus WSeTe [Fig. \ref{fig3:LDOSs}(c)] at the same laser parameters selected in Fig. \ref{fig2:Spin-band}. For 1T$^\prime$-WSe$_2$, a direct TPT from $C$ = 0 to $C$ = +2 occurs at $A_0$ = 65 V/$c$ [the third panel of Fig. \ref{fig3:LDOSs}(a)]. The QAH gap further increases as the laser amplitude rises. A nonzero Berry curvature is observed around the two valleys with positive chirality [see Fig. \ref{fig3:LDOSs}(b), by selecting $A$ = 75 V/$c$ as a representative], indicating a normal QAH state without valley polarization.

	Furthermore, in the valley-polarized case (Janus WSeTe), the imposition of valley polarization leads to an additional Chern number, $C$ = +1, as the light intensity increases [Fig. \ref{fig3:LDOSs}(c)]. Figure \ref{fig3:LDOSs}(d) illustrates the Berry curvature distribution across the 2D BZ of Janus WSeTe ($A_0$ = 70 V/$c$). At valley $K$, a red-colored zone of positive chirality persists, while at valley $K'$, a small positive-chirality zone emerges compared to the previous QVH state before the first TPT point ($A_0$ = 40 V/$c$) [see Figs. S6(b) and S6(c) \cite{supplementary}]. This positive-chirality zone coexists with a left-located negative-chirality zone, indicating a total Chern number of zero contributed by valley $K'$. Consequently, valley polarization becomes evident, leading to the manifestation of this QAH state as a vp-QAH state, expressed as $C_v\ =\ C_K\ -\ C_{K^\prime}\ =\ +1$, underscoring the intricate interplay between valley dynamics and the topological properties of the material. 
	
	% Fig4
	\begin{figure*}
		\centering
		\includegraphics[width=0.85\linewidth]{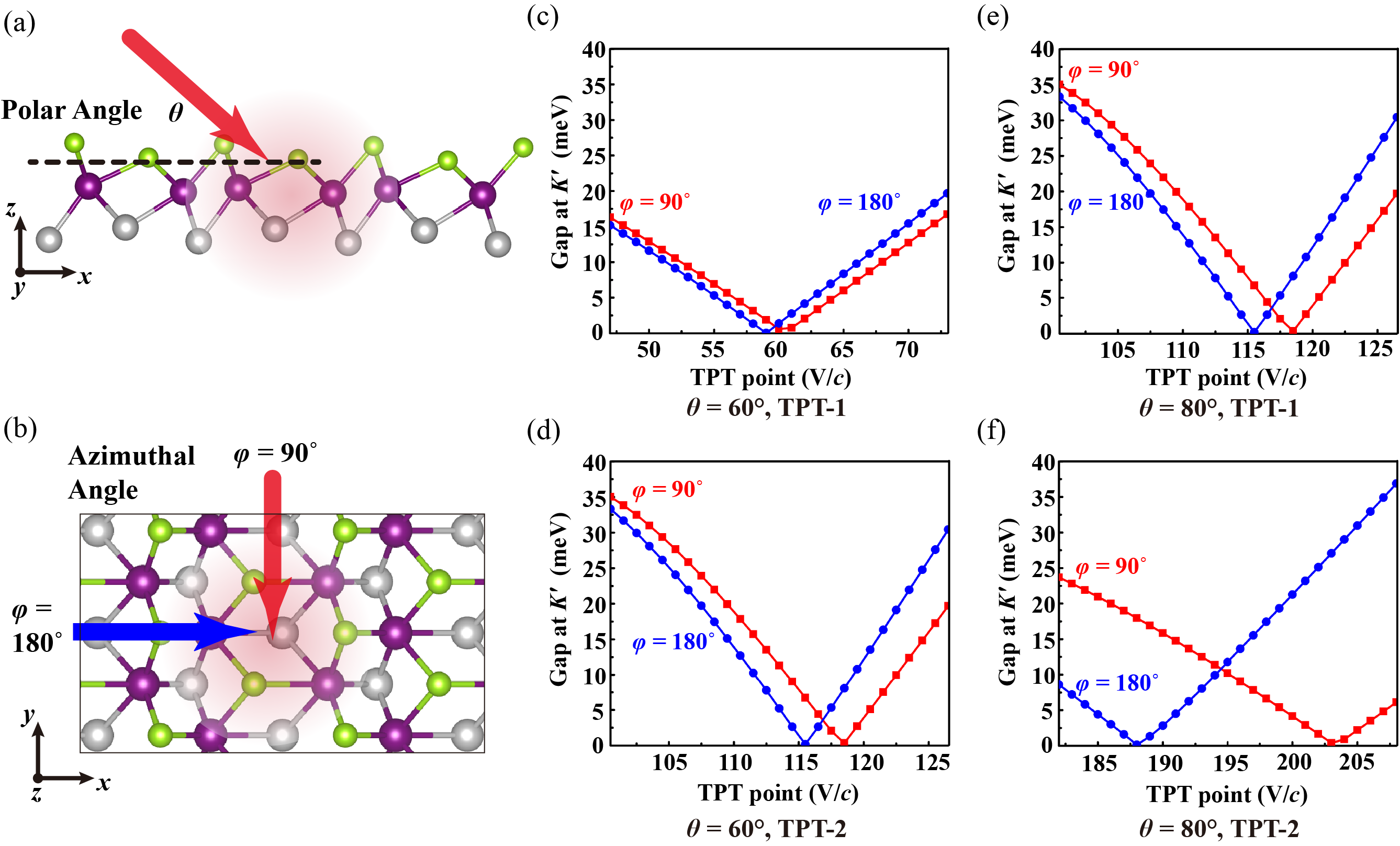}
		\caption{\textbf{In-plane anisotropic response of TPTs based on Janus WSeTe, with the light frequency set to $\hbar\omega$ = 1.0 eV.} (a) The illustration of the incident polar angle ($\theta$) is shown in the side view of Janus WSeTe, with the red arrow indicating the direction of the incident light. (b) presents a similar view, but from the top perspective of Janus WSeTe, where it depicts the azimuthal angle ($\varphi$). The blue and red arrows represent the incident light at azimuthal angles of 90$^{\circ}$ and 180$^{\circ}$, respectively. The black frame envelopes a 2$\times$2 in-plane supercell of Janus WSeTe. Panels (c) and (d) show the evolution of the gap at the $K'$ valley around the first and second TPT points, respectively, as a function of light intensity. In these panels, the red and blue lines correspond to azimuthal angles of 90$^{\circ}$ and 180$^{\circ}$, with $\theta$ set to 60$^{\circ}$. Panels (e) and (f) are similar to (c) and (d), except that the polar angle is set to 80$^{\circ}$.}
		\label{fig4:Anisotropic}
	\end{figure*}
	
	After the second TPT point ($A_0$ = 83 V/$c$) for Janus WSeTe, the negative-chirality zone around valley $K'$ completely collapses, with only positive-chirality zone broadening to the similar size compared to that around valley $K$ [see Fig. S6(e) \cite{supplementary}]. As a result, the vp-QAH state transitions into a normal QAH state, characterized by $C_v\ =\ C_K\ -\ C_{K^\prime}\ =\ 0$ and $C$ = +2 . However, the gap values of the two valleys remain distinct, with the smaller gap determining the final global gap.

	When the chirality of the light is switched to L-CPL, the signs of the total Chern number ($C$), the valley-polarized Chern number ($C_v$), and the valleys associated with the TPT are completely inverted. The LDOS patterns and Berry curvature distributions under this condition are illustrated in Fig. S7 \cite{supplementary}.

	\subsection{Angle-dependent Anisotropic Response}
	
	Notably, the oblique incidence of light introduces an additional degree of freedom for manipulating valley topology. The previously discussed conditions correspond to vertical incidence, where the light wave vector is expressed as: $\textbf{\textit{A}}\ =\ A_0\ (\mathrm{cos}\omega\textbf{\textit{t}},\ \ \mathbf{i}\mathrm{sin}\omega\textbf{\textit{t}},\ 0)$. In the case of oblique incidence, two additional parameters come into play: the polar angle $\theta$, and the azimuthal angle $\varphi$. Consequently, the light vector transforms to: $\textbf{\textit{A}}\ =\ A_0\ (\mathrm{cos}\omega\textbf{\textit{t}}\mathrm{cos}\varphi-\mathbf{i}\mathrm{sin}\omega\textbf{\textit{t}}\mathrm{cos}\theta \mathrm{sin}\varphi,\ \ \mathbf{i}\mathrm{sin}\omega\textbf{\textit{t}}\mathrm{cos}\theta \mathrm{cos}\varphi+\mathrm{cos}\omega\textbf{\textit{t}}\mathrm{sin}\varphi,\ \mathbf{i}\mathrm{sin}\omega\textbf{\textit{t}}\mathrm{sin}\theta)$. When $\theta$ = 0$^{\circ}$ (i.e., vertical incidence), the azimuthal angle $\varphi$ has no effect, and $\textbf{\textit{A}}$ remains unchanged. As the value of $\theta$ increases, more in-plane components of the light are generated, enhancing the tunability of the potential in-plane anisotropic response within the system.
	
	Figure \ref{fig4:Anisotropic}(a) illustrates the incident polar angle ($\theta$), with the side view of Janus WSeTe shown as the background. Similarly, the azimuthal angle ($\varphi$) is depicted in Fig. \ref{fig4:Anisotropic}(b), using the top view of Janus WSeTe. The red and blue arrows represent light with the incident azimuthal angles chosen as 90$^{\circ}$ and 180$^{\circ}$, respectively. Here, we use Janus WSeTe as a template, where both TPT points exhibit angle-dependent anisotropy. Figures \ref{fig4:Anisotropic}(c) and \ref{fig4:Anisotropic}(d) sequentially depict the gap evolution at the $K'$ valley for the first and second TPT points, with the polar angle set to 60$^{\circ}$. The color of the polygonal lines is inherited from the arrows shown in Fig. \ref{fig4:Anisotropic}(b). A staggered nature (with a difference of 1.0 V/$c$) emerges between the red line ($\varphi$ = 90$^{\circ}$) and the blue line ($\varphi$ = 180$^{\circ}$), confirming that the TPT occurs slightly later in the former case. This phenomenon reveals the anisotropic response based on the azimuthal angle. Moreover, the second TPT point also exhibits staggered behavior, but with even larger differences: 118.5 V/$c$ for $\varphi$ = 90$^{\circ}$ and 115.5 V/$c$ for $\varphi$ = 180$^{\circ}$, respectively [Fig. \ref{fig4:Anisotropic}(d)].

	As $\theta$ increases, the in-plane optical component becomes more pronounced, thereby enhancing the anisotropy. Figures \ref{fig4:Anisotropic}(e) and \ref{fig4:Anisotropic}(f) sequentially illustrate the gap evolution around the two TPT points at the $K'$
	valley, with $\theta$ set to 80$^{\circ}$. The first TPT point exhibits values of 118.5 V/$c$ for $\varphi$ = 90$^{\circ}$ and 115.5 V/$c$ for $\varphi$ = 180$^{\circ}$. Moreover, the second point demonstrates an even greater degree of staggered behavior, yielding 203.0 V/$c$ for $\varphi$ = 90$^{\circ}$ and 188.0 V/$c$ for $\varphi$ = 180$^{\circ}$. These results provide insight into how light at a large angle of oblique incidence can effectively detect the in-plane anisotropic characteristics of the valleys, thereby indirectly revealing their symmetries.

	The microscopic origin of the aforementioned anisotropic Floquet response can be understood through the coupling between spin-polarized $d$ orbitals of W and $p$ orbitals of the chalcogen atoms, where the real-space distribution of charge densities varies along the in-plane directions. Through Peierls substitution, the periodic optical field induces distinct band reshaping along different reciprocal paths, as confirmed by the computational results presented above. These features can be experimentally accessed using angle-resolved photoemission spectroscopy (ARPES) or incident-angle-dependent Hall effect measurements.

	While the anisotropic response is primarily influenced by the in-plane component of the light, the vertical component remains crucial. When the polar angle reaches 90$^{\circ}$, indicating parallel incidence: no phase transition can be induced within the laser parameter range utilized in this study (see Fig. S8 of the Supplementary Materials \cite{supplementary}). These two components are non-commutative and collectively contribute to the formation of this anisotropic response.

		% Fig5
	\begin{figure*}
		\centering
		\includegraphics[width=0.93\linewidth]{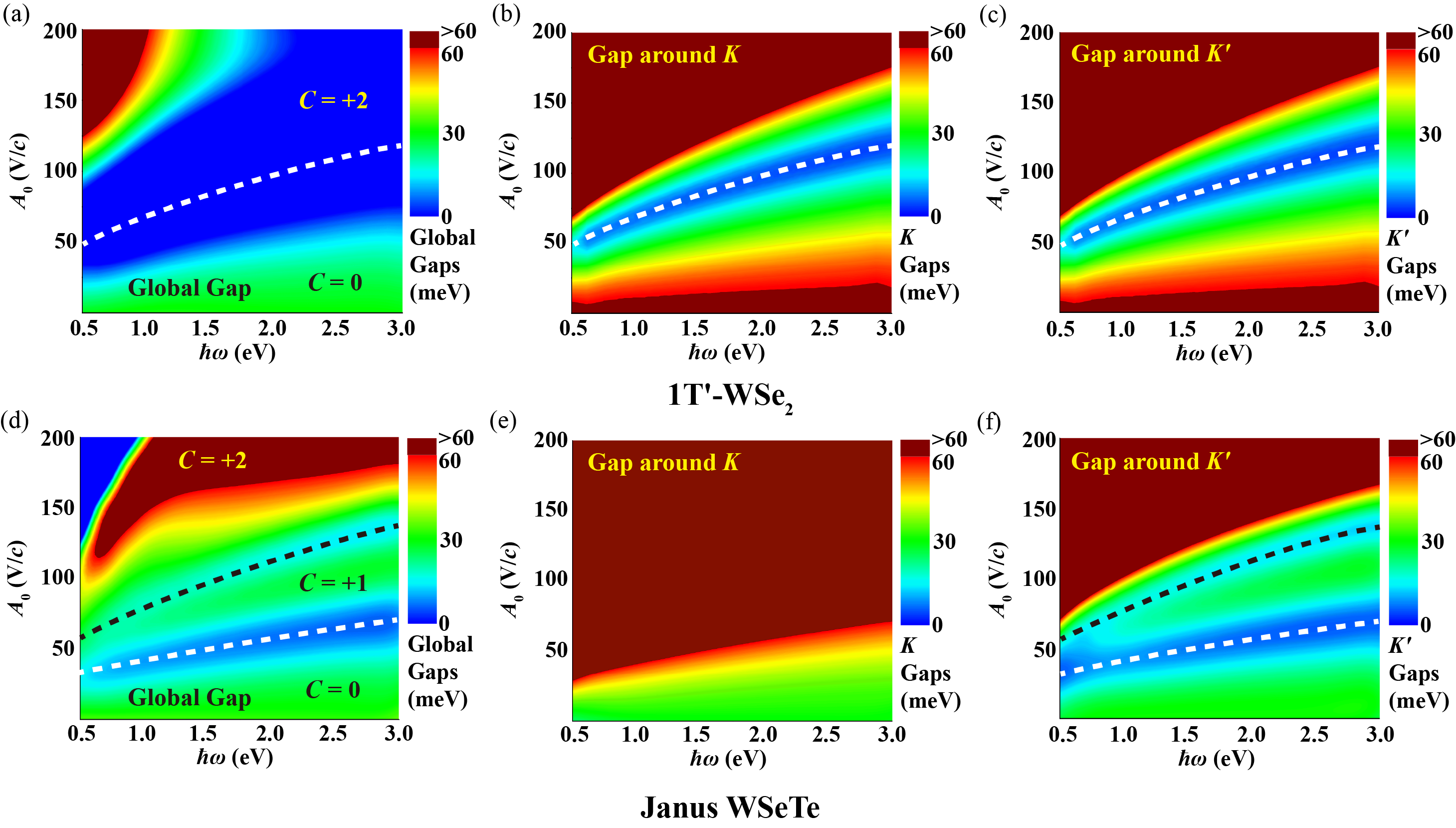}
		\caption{\textbf{Contour distributions of global gaps and valley gaps based on various laser parameters.} Panels (a)-(c) are related to 1T$^\prime$-WSe$_2$, with subfigure (a) providing the distribution of the global gap, while subfigures (b) and (c) depict the gaps around the $K$ and $K'$ valleys, respectively. Panels (d)-(f) correspond to Janus WSeTe and are analogous to (a)-(c). In each contour distribution, the color gradient transitions from blue to green, yellow, and then to red, indicating an increase in gap values, meanwhile the maroon region represents an ultra-large gap exceeding 60 meV. The laser parameters vary along the $x$-axis (light frequency) and $y$-axis (light intensity). Additionally, the white dashed curve indicates the first TPT point, while the black dashed curve signifies the second TPT point, which is present only in Janus WSeTe.}
		\label{fig5:Contours}
	\end{figure*}

	\subsection{Laser-Parameter Phase Diagram }
	
	To select laser parameters that achieve large global gaps and various Chern numbers, we typically conduct a contour distribution analysis of the global gap, as well as the local gap distributions at the $K$ and $K'$ points, as a function of laser amplitude and frequency. 1T$^\prime$-WSe$_2$ features only one TPT point, where the laser amplitude is related to the laser frequency by a square root relationship [Fig. \ref{fig5:Contours}(a)]. However, a large metallic zone (depicted in dark blue) is observed surrounding the TPT curve, primarily due to the gap misalignment between the $K$ ($K'$) valley and the $\Gamma$ point, as shown in the LDOS patterns in Fig. \ref{fig3:LDOSs}(a).
	
	Increasing the laser amplitude slightly can restore the global gap within the Chern-insulating region, with some areas achieving an ultra-large gap ($\geq$ 30 meV), indicated by the green, yellow, red, and maroon zones in the top-left corner of the plot. Notably, 1T$^\prime$-WSe$_2$ still supports a QAH state at temperatures beyond room temperature. Figures \ref{fig5:Contours}(b) and \ref{fig5:Contours}(c) present the local gap distributions at the $K$ and $K'$ valleys, respectively, revealing identical patterns that are ensured by the symmetry of the system. In the absence of the $\Gamma$-point valence bands, the metallic zone disappears, and the tiny-gap region diminishes near the TPT curve.

	Fortunately, the monolayer Janus WSeTe overcomes the metallic behavior without the influence of the $\Gamma$-point valence bands, as evidenced by the LDOS pattern shown in Fig. \ref{fig3:LDOSs}(c). Consequently, in the global gap distributions presented in Fig. \ref{fig5:Contours}(d), the metallic region is largely absent, relegated to a small area in the top-left corner of the phase diagram. The two TPT curves are situated within very narrow blue zones, indicating a sharp closure and reopening of the band gap at both TPT points.

	The valley polarity in Janus WSeTe creates a distinct divergence between the two valleys. The $K$ valley [Fig. \ref{fig5:Contours}(e)] exhibits no phase transition and demonstrates a monotonic enhancement of the gap, while the $K'$ valley [Fig. \ref{fig5:Contours}(f)] undergoes a two-stage phase transition, which ultimately determines the final global gap. A wide range of ultra-large global gaps can be achieved in the final phase with $C$ = +2. However, only up to 30 meV of global gap is observed in the intermediate vp-QAH phase with $C$ = +1, which is still sufficient for room-temperature valley-polarized Chern insulators. These results demonstrate that Janus WSeTe sustains large global gaps ($\textgreater$ 30 meV) across multiple topological regimes, confirming its feasibility for room-temperature Floquet valleytronics.

	\section{Discussion}
	
	Considering that six cycles of a periodic optical field can fabricate Floquet replica bands with zero-order band reshaping \cite{qiu2018ultrafast}, the maximum light frequency ($\hbar\omega$ = 3.0 eV) and laser amplitude ($A_0$ = 200 V/$c$) employed in this study yield a maximum fluence of approximately 0.010 J/cm$^2$. In comparison to the maximum fluence of 0.013 J/cm$^2$ used in group-IV Xenes \cite{li2025light}, this approach appears to be gentler. Unfortunately, there is currently no experimental evidence regarding the thermal damage threshold for 1T$^\prime$-WSe$_2$ and Janus WSeTe. A similar work of the photovoltaic effect based on 1T$^\prime$ Janus transition metal dichalcogenides has adopted the light intensity up to 0.25 Å$^{-1}$ ($\sim$ 490 V/$c$) \cite{zhao2025prediction}, much stronger than those employed in this work. To further investigate, we conduct a bond-energy assessment of these two systems in comparison with monolayer Xenes and make additional predictions. The formation energies for 1T$^\prime$-WSe$_2$ and Janus WSeTe monolayers are --3.93 eV/bond and --3.67 eV/bond, respectively. These values are closer to that of stanene (--3.79 eV/bond) and represent about 73\% of the formation energy of germanene (--5.03 eV/bond) and 59\% of that of silicene (--6.25 eV/bond). Taking into account the thermal damage threshold for silicene and germanene mentioned in Ref. \cite{li2025light}, which is 0.050 J/cm$^2$ \cite{sokolowski2001femtosecond}, the maximum fluence of 0.010 J/cm$^2$ used in this study can be considered safe for the systems under investigation.
	
	\section{Conclusion}
	
	In conclusion, going beyond conventional triangular or hexagonal lattices, we focus on a less symmetric, orthorhombic building block monolayer 1T$^\prime$-WSe$_2$-family materials and its Janus derivative WSeTe as anisotropic platforms for exploring Floquet-engineered valley topology. Our findings systematically reveal the entire evolution of valley-polarized topology, featuring a two-stage TPT process that shows a transformation of the Chern number from 0 to $\pm$1, and finally to $\pm$2 across a pair of valleys. This valley-polarized topology is manifested by Janus WSeTe, where the two stages of TPT occur at a single valley but are sequentially driven by the two spin channels. In the intermediate Chern insulating state, valley polarity persists, as evidenced by Berry curvature distributions, leading to a vp-QAH state with $C_v$ = +1. By appropriately adjusting the laser parameters, this vp-QAH state demonstrates significant potential for operation at room temperatures.

	Moreover, the anisotropic placement of the valleys within the $\textbf{\textit{a}*}$-$\textbf{\textit{b}*}$ plane gives rise to an angle-dependent Floquet response under oblique light incidence. This anisotropy enables directional control of valley topology through the incident light orientation, reflecting the intimate coupling between crystal symmetry and Floquet modulation. Overall, our study establishes anisotropic 1T$^\prime$-WSe$_2$ and Janus WSeTe monolayers as promising platforms for light-controllable topological and valleytronic functionalities.
	
	$Note$ $added.$  Recently, we became aware of a similar study on 1T$^\prime$ Janus transition metal dichalcogenides, which also exhibits a two-stage TPT as the intensity of CPL increases, transitioning through Chern numbers of 0, $\pm$1 and $\pm$2 as the CPL intensity enhances \cite{zhao2025prediction}. Compared to this work, our manuscript provides a more detailed physical analysis of valley polarity engineered by the Floquet-Bloch theorem. We also present a tunable pathway between QSH, QAH, QVH, and vp-QAH states, manipulated by structural symmetry and light intensity. Furthermore, the strong in-plane anisotropy of these 1T$^\prime$ transition metal dichalcogenides enables the observation of anisotropic responses in TPTs driven by varying incident azimuthal angles of light.

	\begin{acknowledgments}
		We thank Dr. Hui Zhou and Dr. Xiyu Hong for helpful discussions, and also thank Dr. Hang Liu for technical supports. The numerical calculations have been done on the supercomputing system in the Huairou Materials Genome Platform. This work is supported by National Natural Science Foundation of China (Grant Nos. 12025407, 12450401,  12104072 and 12304536), National Key Research and Development Program of China (Grant No. 2021YFA1400201), and Chinese Academy of Sciences (Grant Nos. YSBR-047 and XDB33030100).
	\end{acknowledgments}
	
	\newpage
	\nocite{*}
%	\bibliography{Main_text}

\end{document}